\documentclass[prl,amsmath,aps,twocolumn,superscriptaddress,letterpaper]{revtex4}
\usepackage{xspace,units,subfigure}
\usepackage{float}
\usepackage{amsmath}
\usepackage{amssymb}
\usepackage[normalem]{ulem}
\usepackage[pdftex]{graphicx}

\begin{document}


\title{The Unequal Twins - Probability Distributions Aren't Everything}

\author{Yasmine Meroz}
\email{merozyas@post.tau.ac.il}
\affiliation{School of Chemistry, Raymond \& Beverly Sackler Faculty of 
Exact Sciences, Tel-Aviv University, Tel Aviv 69978, Israel}
\author{Igor M. Sokolov}
\affiliation{Institut f\"{u}r Physik, Humboldt-Universit\"{a}t zu Berlin, 
Newtonstrasse 15, D-12489 Berlin, Germany }
\author{Joseph Klafter}
\affiliation{School of Chemistry, Raymond \& Beverly Sackler Faculty of 
Exact Sciences, Tel-Aviv University, Tel Aviv 69978, Israel}

\begin{abstract}
It is the common lore to assume that knowing the equation for the probability distribution
function (PDF) of a stochastic model as a function of time tells the whole picture defining 
all other characteristics of the model. We show that this is not the case by comparing two 
exactly solvable models of anomalous diffusion due to geometric constraints: 
The comb model and the random walk on a random walk (RWRW). We show that though the
two models have exactly the same PDFs, they differ in other respects, 
like their first passage time (FPT) distributions, their autocorrelation functions 
and their aging properties.
\end{abstract}

\maketitle

\begin{figure}[h]
\centerline{
  \mbox{\includegraphics[width=0.25\textwidth,]{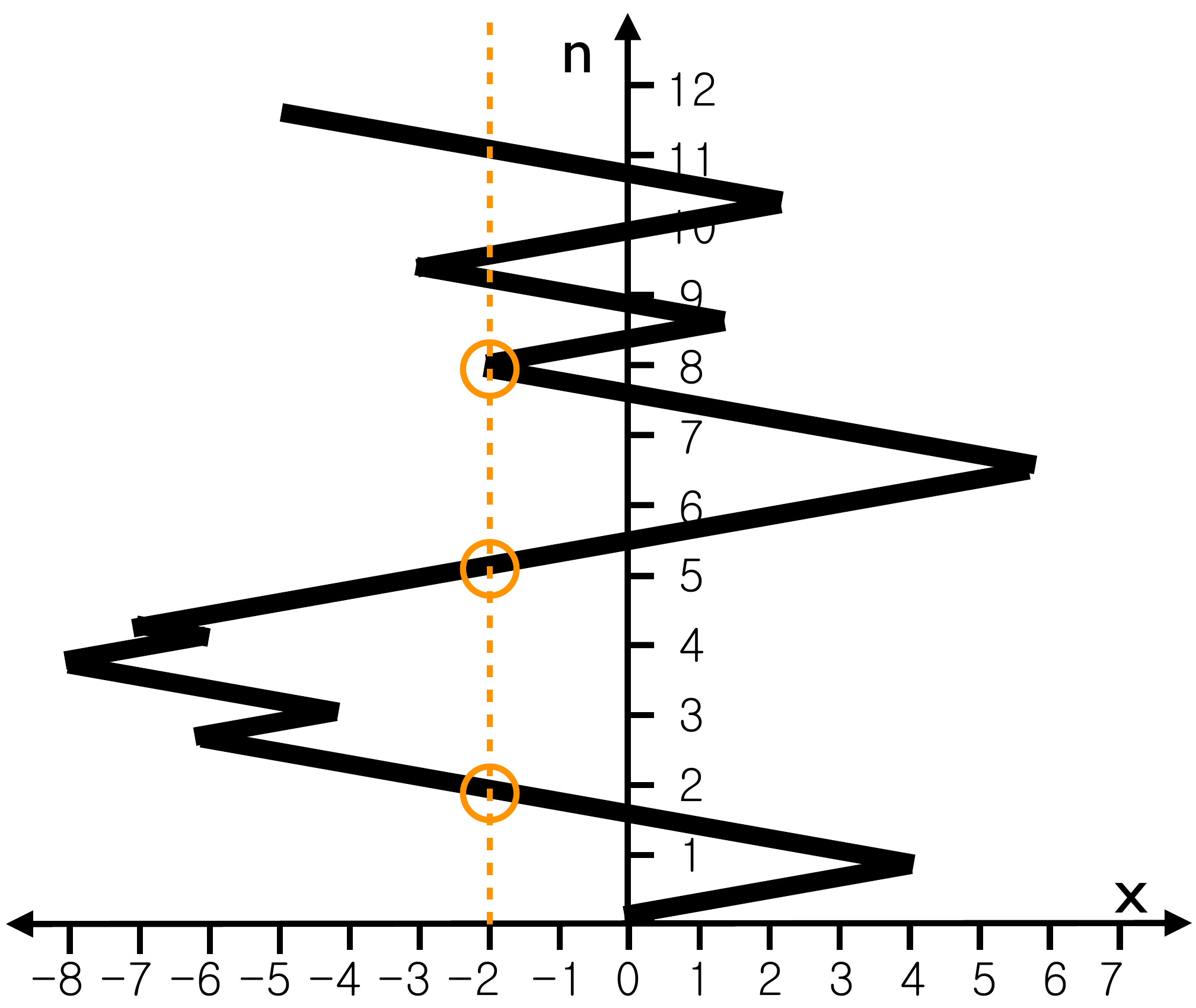}}
  \mbox{\includegraphics[width=0.25\textwidth,]{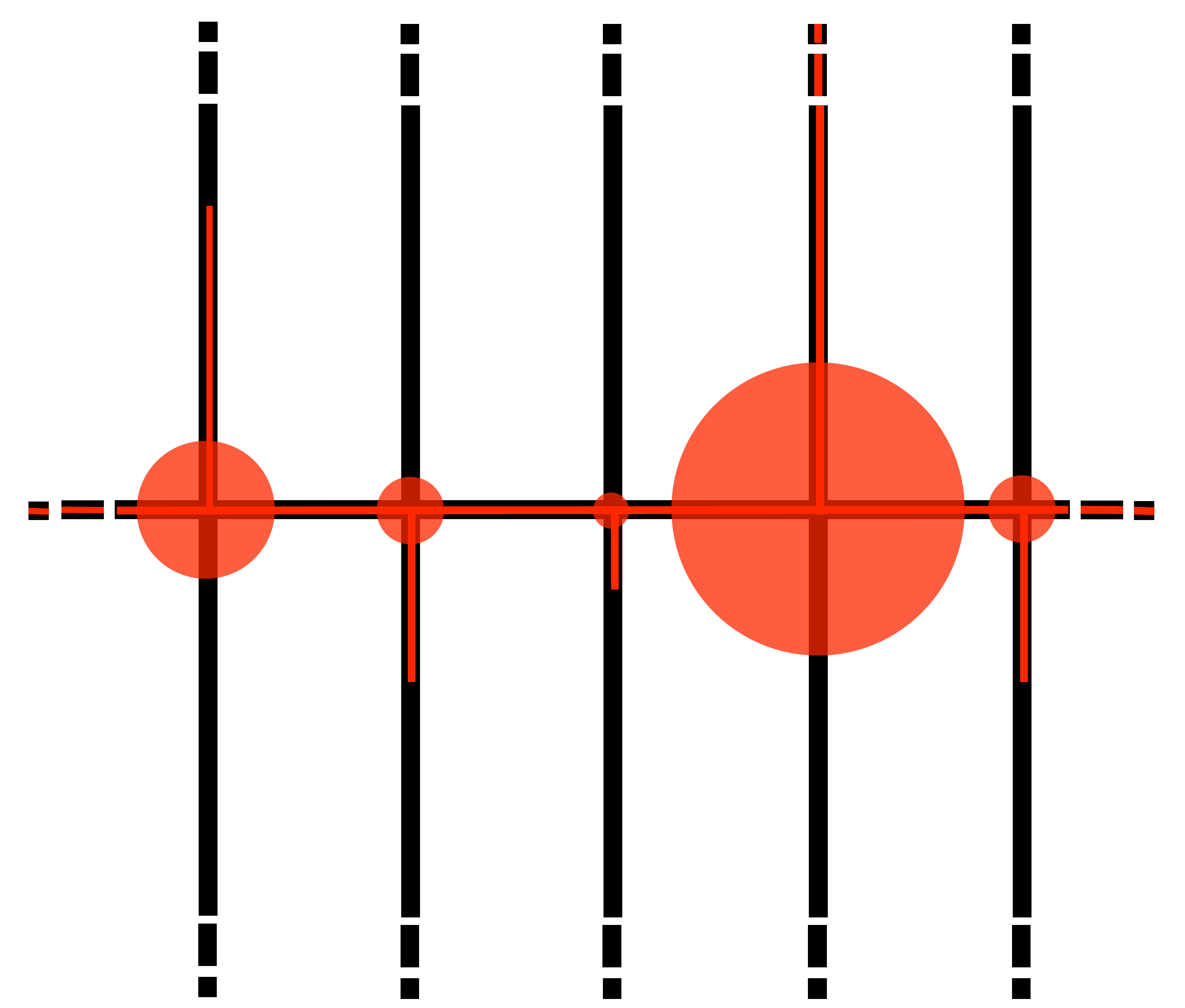}}
}
\caption{(color online) Left: RWRW model. The trajectory 
of a random walk (acting as a track on which a RWer may move) is displayed in black. Circles mark
the position $x=-2$ at different 'chemical distances' along the
track. Right: A RW on a symmetric comb structure. The RW trajectory is marked in a lighter 
shade (red). The projection of the trajectory on the $x$-axis, or backbone, leads to waiting times caused by the RW's sojourn in
a tooth (depicted by  
circles of sizes representing the appropriate waiting time), thus a CTRW.}
\label{fig:models}
\end{figure}

One often thinks that perfect knowledge of the PDF as a function of time
(achieved by better experimental techniques or by data mining) completely 
determines the underlying stochastic model. In what follows we 
illustrate that this is not the case by 
revisiting two exactly solvable models often 
invoked when explaining anomalous diffusion in labyrinthine structures 
of percolation type: the comb model introduced in~\cite{Weiss1986}
(which mimics trapping in the dangling ends) 
and the RWRW  introduced in \cite{Kehr1982} (which mimics the tortuosity of the chemical path). 
Both are visualized in Fig.~\ref{fig:models}. The RWRW model was
introduced as a simplified (quenched) version of the single-file diffusion.
In polymer physics, the RWRW corresponds to one of the
regimes of reptation~\cite{Doi1988}, namely, to the
motion of the chain as a whole along its primitive path.
These facts make evident its relation to other models of diffusion 
under time-dependent constraints, which are often described using a
fractional Brownian motion approach. Although both models
are known for around quarter of a century and were quite carefully
investigated, the two were never confronted. Doing so is
however quite instructive. Thus in the long time limit both models,
the comb (of the CTRW class~\cite{Weiss1986}) and the RWRW (in a class of itself, but a
close relative of the single file diffusion) show \textit{exactly the same} PDF which can be
described by the corresponding fractional diffusion equation~\cite{Metzler2000}. However,
the models differ in any other respect: the comb model exhibits aging
while the RWRW does not, the distribution of FPTs
to a given site are different and also the auto-correlation functions differ.

Note that both the comb and RWRW models stem from Markovian random
walk (RW) models on the corresponding geometries, i.e., the
probabilities of steps from a site $(i,j)$
on the underlying track depend only on the position $(i,j)$.
In both cases the RWs have independent but not identically
distributed increments. We are interested in the projection of the RW onto the $x$-axis.
The fact that the same value of $x$ may correspond to different values of $y$
in the comb model, or to different positions along the contour length
$n$ in the RWRW model (see Fig.~\ref{fig:models}), and that the further motion depends explicitly
on this $y$ or $n$,
introduces the memory on the previous positions. Introducing the memory is the cost we pay for
neglecting the {\it irrelevant} coordinate ($y$ or $n$), the dependence on which
however does not disappear. The kind of non-Markovian behavior of the reduced
models is strongly different: The diffusion on the comb falls into a
semi-Markovian CTRW class, the one on the RW track is an
anti-persistent RW, a close relative (but not a member of the
class) of fractional Brownian motions.


In the CTRW as represented by the comb model the probability density of
being at position $x$ after $n$ time steps is given by the subordination expression
\begin{equation}\label{eq:subordination}
p(x,t) =\sum_{n=0}^{\infty} p(x,n) \chi(n,t) \simeq \int_0^{\infty}{p(x,
n)\chi(n,t) \mathrm{d}n}
\end{equation}
where $p(x,n)$ gives the probability of reaching $x$ after $n$ steps and $\chi(n,t)$
is the probability to make $n$ steps within time $t$.
Here $t$ is measured in the units of time steps $\tau=1$.
Note that for $t \gg \tau$ the typical number of steps $n$ is large and
can be considered as a continuous variable, turning the sum in
Eq.~(\ref{eq:subordination}) into an integral over $n$.
The waiting time distribution along the comb backbone is given by the
return time probability of a RW on a tooth to the backbone, which is
approximately $\psi(t) \simeq (2 \pi)^{-1/2} t^{-3/2}$ \cite{Chandrasekhar1943}.
In the Laplace domain, $\chi(n,t)$ and $\psi(t)$ are related via~\cite{KlafterSokolov}:
\begin{equation}\label{eq:chin_laplace}
\chi(n,s) = s^{-1}[1-\psi(s)] \psi^n(s),
\end{equation}
where $f(s)$ denotes the Laplace transform of a function $f(t)$.
Substituting $\psi(s) \simeq 1 - \sqrt{2} s^{1/2}$ into
Eq.~(\ref{eq:chin_laplace}) and performing the inverse Laplace transform
results in: 
\begin{equation}\label{eq:chin}
\chi(n,t) \simeq \frac{ \sqrt{2}}{\sqrt{\pi t}} \exp\left(-\frac{n^2}{2t} \right),
\end{equation}
a Gaussian distribution restricted to the non-negative half-line. 
Substituting Eq.~(\ref{eq:chin}) and $p(x,n)$ (which tends
to a Gaussian for large $n$ for a regular RW on the
backbone) into the continuous limit of Eq.~(\ref{eq:subordination}), 
we get:
\begin{equation}\label{eq:pdfCOMB}
p(x,t) = \int_0^{\infty} \frac{1}{\sqrt{2 \pi n}} \exp \left(-\frac{x^2}{2 n} \right)
\frac{\sqrt{2}}{\sqrt{ \pi n}} \exp \left(-\frac{n^2}{2 t} \right) \mathrm{d}n.
\end{equation}

\begin{figure}[t]
\begin{centering}
\includegraphics[width=0.45\textwidth,]{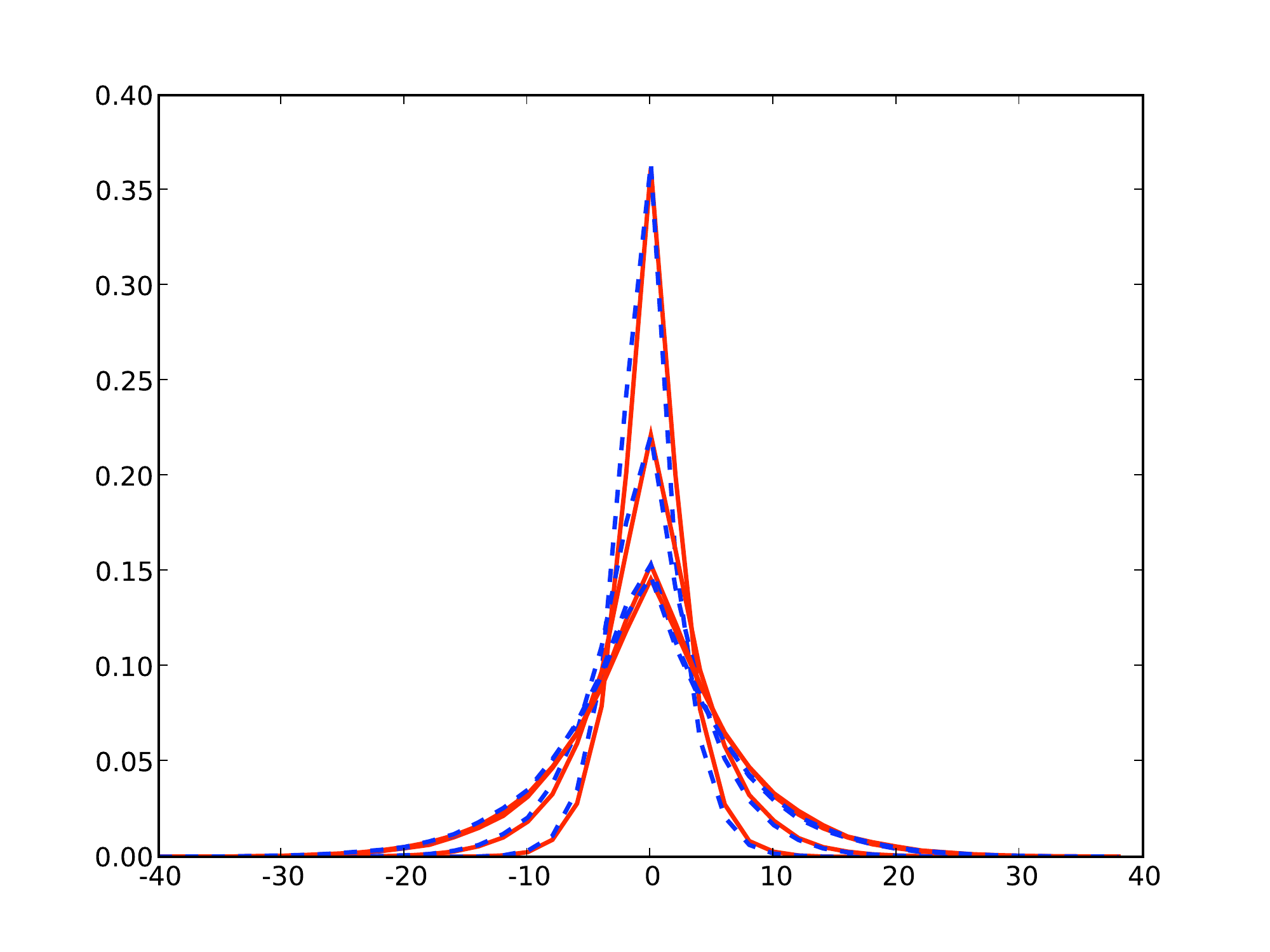}
\par\end{centering}
\caption{(color online) PDFs of RWRW (smooth red line) vs.  RW on a
comb (dashed blue line) for a different number of time steps: (a) t=100,
(b) t=1000, (c) t=5000, (d) t=6500. It can be clearly seen that the
PDFs of the two models coincide.
\label{fig:PDFs}}
\end{figure}

Let us now calculate the PDF of the particles' displacement in the RWRW
model, pertinent to many realizations of walks and starting points.
Let us consider a RW on a RW starting at $x=0$
and consider its position after $t \gg 1$ time steps
in a continuous approximation when the displacement $n$
of the RWer \textit{along the trajectory} of the RW
is well-approximated by a Gaussian:
\begin{equation}\label{eq:plRWRW}
\kappa(n,t) \simeq \frac{1}{\sqrt{2 \pi t}} \exp \left(-\frac{n^2}{2t} \right).
\end{equation}
The projection on $x$ of this displacement along the trajectory is
again given by a Gaussian depending on the displacement $|n|$:
\begin{equation}\label{eq:pxRWRW}
p(x,n) \simeq \frac{1}{\sqrt{2 \pi |n|}} \exp \left(-\frac{x^2}{2|n|} \right).
\end{equation}
The displacement in $x$ as a function of $t$ is expressed by:
\begin{equation}\label{eq:pdfRWRW}
p(x,t) = \int_{-\infty}^{\infty} p(x,n) \kappa(n,t) \mathrm{d}n.
\end{equation}
Substituting Eqs.~(\ref{eq:plRWRW}) and~(\ref{eq:pxRWRW}) into
Eq.~(\ref{eq:pdfRWRW}) results in 
\textit{exactly} the same expression as Eq.~(\ref{eq:pdfCOMB}).
This is numerically verified in Fig.~\ref{fig:PDFs}. The models are essentially twins!

\begin{figure}[h]
\begin{centering}
\includegraphics[width=0.45\textwidth,]{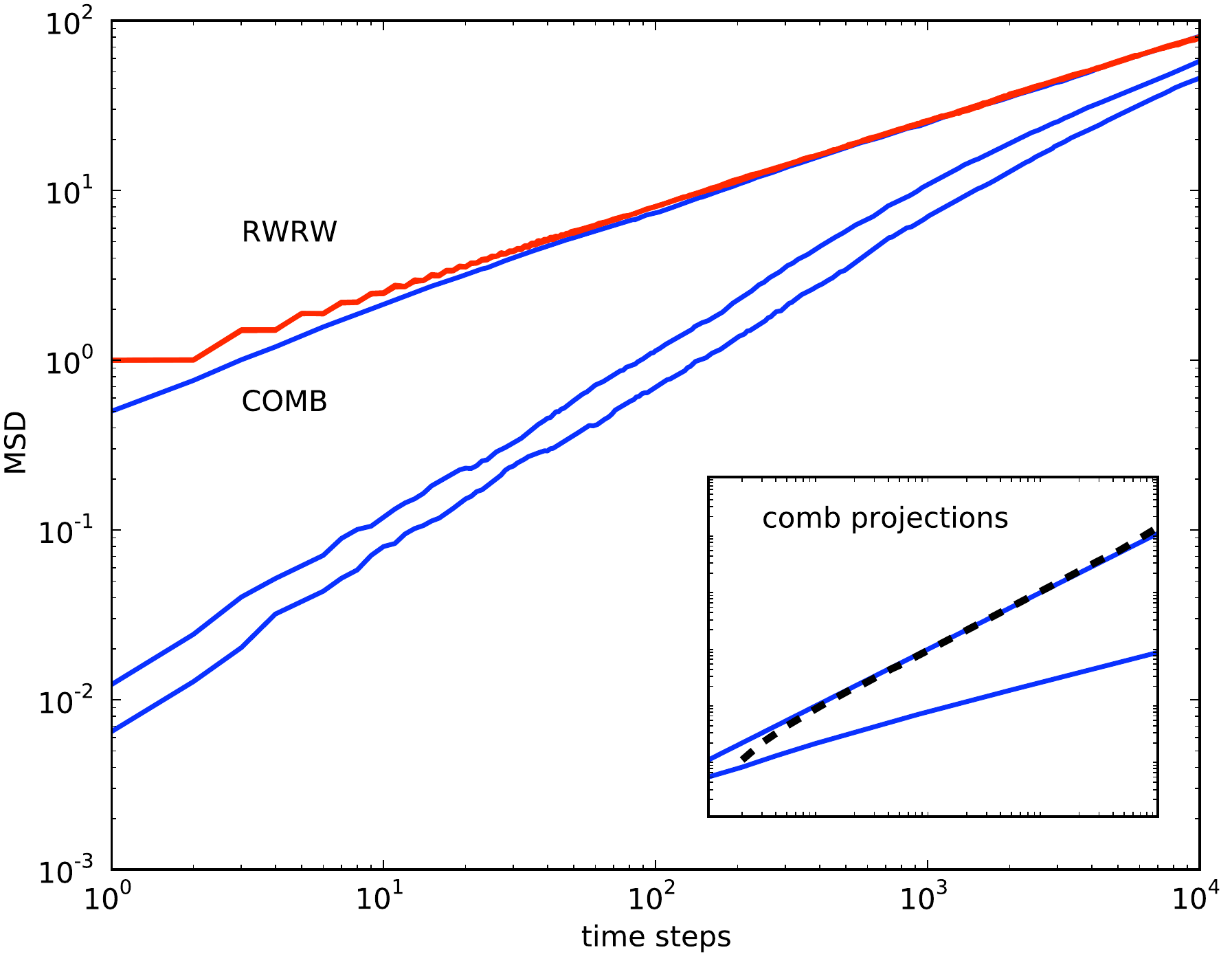}
\par\end{centering}
\caption{(color online) MSDs in a comb and in a RWRW
for different time lags after the beginning of the RW ($\tau=0,1000, 3000$). The
MSDs for different time lags for the RWRW overlap (upper, red curve).
The lower (blue) curves represent the MSDs of the comb model. The larger the time lag the longer it takes the MSD to reach its
asymptotic behavior. The inset shows the MSD
of the the 2d comb model averaged over time 
(black dashed line) and over an ensemble (overlapping solid blue),
showing the ergodicity of the two-dimensional motion. The behavior of $\langle x^2\rangle$  
(lower solid blue) is shown for comparison.
\label{fig:MSD_RWRW_COMB}}
\end{figure}

Let us however stress a fundamental difference between the two models.
Eq. (\ref{eq:pdfCOMB}) implies that the time $t$ is
counted from the beginning of the walk starting at $(0,0)$. 
In the course of time the walker typically enters deeper into the dangling ends. If the observation 
starts at a later instant $t>0$, the walker will typically need time to 
return to the $x$-axis to take a step in a relevant direction. Therefore 
the waiting time PDF for the first step after $t>0$ 
is different from $\psi(t)$~\cite{KlafterSokolov}, leading to a different $\chi(n,t)$: 
the diffusion on the comb shows \textit{aging} typical of CTRW~\cite{Sokolov2001, Sokolov2001a}. 
On the contrary, the RWRW model does not age: \textit{the same}
distribution $\kappa(n,t)$ holds, whenever one starts. This property
can be inferred from Fig.~\ref{fig:MSD_RWRW_COMB} showing MSDs for both models 
for observations starting at different times from the beginning
of the RW.

Note that the movements in the $x$ and $y$ directions on the comb are correlated: the more steps 
in the $x$ direction are made, the less time remains for steps in the $y$ direction. 
The total MSD $\langle r^2 \rangle =\langle x^2 \rangle + \langle y^2 \rangle$ is found to be
diffusive, $\langle r^2(t) \rangle \sim n $, \cite{SupMat1}, 
which is numerically verified (inset in Fig.~\ref{fig:MSD_RWRW_COMB}). It is
interesting to note that though the $x$-projection of this RW follows
a non-ergodic subdiffusive process, the 2d picture corresponds to 
ergodic normal diffusion. This emphasizes the importance of
vigilance on experimentalists' side when assuming that a projection
onto a lower dimension gives a representative description of the
process. 

Let us now turn to the properties of the process other than the PDF and calculate the 
FPT distributions for the two models, starting with the comb. 
As in a general CTRW, the FPT for a comb is given by the subordination formula
\begin{equation}\label{eq:subordinationFPT}
f(t) = \sum_{n=0}^{\infty} f(n) \psi_n(t)
\end{equation}
where $f(n)$ is the FPT probability at the $n$-th step in a regular RW
and $\psi_n(t)$ is the probability density to make the $n$-th step
at time instant $t$ (note that the first passage only takes place 
exactly when the step is taken, so that $\psi_n(t)\neq \chi(n,t)$). The FPT of a
regular RW is asymptotically given by the Smirnov distribution: 
\begin{equation}\label{eq:smirnov}
f(n) \simeq \frac{|x|}{\sqrt{2 \pi} n^{3/2}} \exp\left(-\frac{x^2}{2 n}\right).
\end{equation}
In the continuum limit one may approximate the sum by an integral over $n$.
Since in the Laplace domain $\psi_n(s) = \psi^n(s)$ for $s \to 0$ we have
$\psi_n(t) \simeq \exp(n(1-\psi(s))$, resulting in $f(s) \sim  \int_0^\infty f(n) e^{-n \sqrt{2s} } dn = e^{-2|x||s|^{1/4}}$, 
which is the characteristic function of a one-sided
L\'evy law of index $1/4$. Its behavior for $t
\to \infty$ is 
\begin{equation}
f(t) \simeq \frac{-2|x|}{\Gamma(-1/4)} t^{-5/4}.
\end{equation} 
We stress that the same result follows from the solution of the fractional diffusion equation
with an absorbing boundary \cite{KlafterSokolov}.

\begin{figure}[h]
\begin{centering}
\includegraphics[width=0.48\textwidth,]{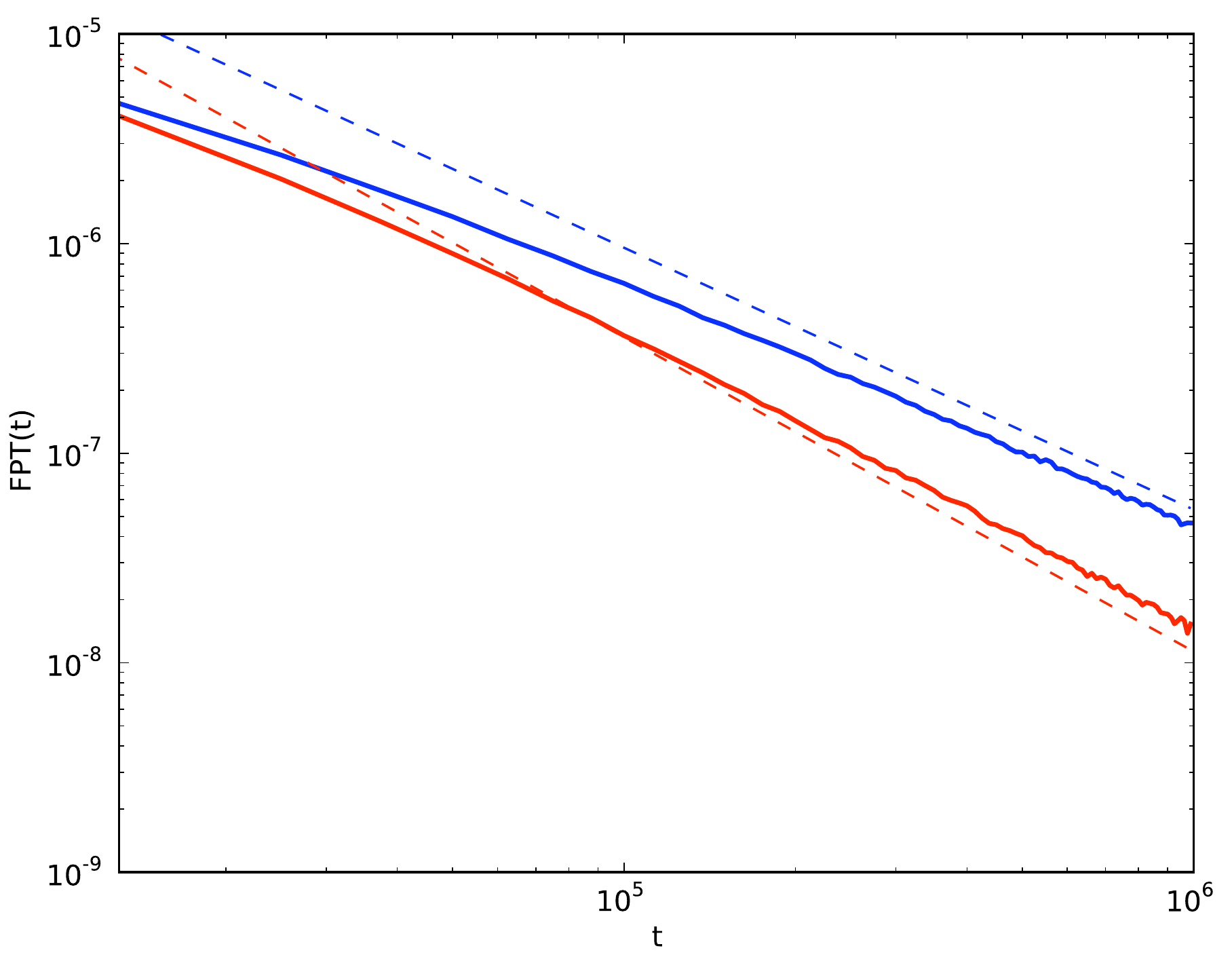}
\par\end{centering}
\caption{(color online) FPT densities for the comb (upper curve, blue) and for the RWRW
(lower curve, red). The dashed lines give the theoretical asympototic behaviours.
\label{fig:FPTs}}
\end{figure}

Let us now calculate the FPT distribution for the RWRW model~\cite{SupMat1}.
We fix the starting point $x=0$ and the final point $x_1$. The RW track on which the
RW then takes place has two strands connecting $x_0$ to the two
points corresponding to $x_1$ (existing due to a recurrence of RWs in 1d). The distances $l_1$ and $l_2$ along
the strands are independent random variables whose
distributions $p(l_1)$ and $p(l_2)$ follow the Smirnov law, Eq.~(\ref{eq:smirnov}).
Therefore the corresponding RW on the track takes place on a finite interval $(-l_1,l_2)$ starting at
$l=0$.
Changing variables we pass to a symmetric interval $(-L/2,L/2)$ of length $L=l_1+l_2$
with the starting point at $y = (l_1-l_2)/2$ inside it.
The survival probability of a particle within this interval with
absorbing boundaries (in the diffusion approximation, $D=1/2$) is given by 
\begin{equation}
\Phi_{1,2}(t) = \sum_{m=0}^{\infty} (-1)^ma_m \exp\left( -\frac{\pi^2 (2m+1)^2}{L^2} Dt\right)
\end{equation}
where $a_m = 4 \cos \left(\frac{\pi (2m+1)y}{L}\right)/\pi (2m+1)$. At long times only the first term ($m=0$) matters, so that
\begin{equation}\label{eq:phi12}
\Phi_{1,2}(t) \approx  \frac{4}{\pi} \cos\left(\frac{\pi y}{2L}\right)
\exp\left( -\frac{\pi^2}{L^2} Dt\right)
\end{equation}
where $Dt=n/2$. Averaging over the position of the
starting point $y$ and over the length of the interval gives:
\begin{equation}\label{eq:phi_int}
\Phi(t) = \iint\limits^{\:\:\:\:\:\:\:\:\infty}_0\Phi_{1,2}(t) p(l_1)p(l_2) \mathrm{d}l_1 \mathrm{d}l_2.
\end{equation}
Substituting Eqs.~(\ref{eq:phi12}) and~(\ref{eq:smirnov}) into
Eq.~(\ref{eq:phi_int}) and changing variables to $\xi=1/(l_1+l_2)$
and $z=l_2/l_1$ we get
\begin{eqnarray}
\Phi(t) = \frac{x^2}{\pi^2}\iint\limits^{\:\:\:\:\:\:\:\:\infty}_0
\frac{1+z}{z^{3/2}} \cos \left(\frac{\pi}{2}\frac{1-z}{1+z} \right) \times \nonumber \\
\exp {\left( -\pi^2 Dt \xi^2 -\frac{x^2(1+z)^2}{4z} \xi \right)} \mathrm{d}\xi \mathrm{d}z.
\label{SurProAv}
\end{eqnarray} 
Evaluating Eq.(\ref{SurProAv})
for $t\rightarrow \infty$
we get~\cite{SupMat1}: $\Phi(t) \simeq C \frac{x^2}{\sqrt{Dt} }$
where $ C \approx 0.64 $ is a number constant. 
Thus the survival probability follows the pattern of a simple
1d RW with the FPT density
\begin{equation}
f(t) \simeq \frac{C x^2}{2\sqrt{D}} t^{-3/2},
\end{equation} 
in agreement with~\cite{Meroz2011}, and not the slower decay of CTRW of $f(t) \propto
t^{-5/4}$. These results are numerically verified and shown in
Fig.~\ref{fig:FPTs}. The anomaly of diffusion however shows up in the
$x$-dependence of the characteristic decay time $t_c$ (say the median
of the survival probability), which scales as $t_c \propto x^4$, exactly like in the case
of the comb, and not as $t_c \propto x^2$ like for the simple diffusion.

Another fundamental intrinsic characteristic is the auto-correlation function. 
Let $s_i$ be the $i$-th step of the walker, so that  
$x_n = \sum_{i=0}^t s_i$. The MSD is therefore:
\begin{equation}
\left \langle x^2_n \right \rangle = \left \langle\sum_{i,j=0}^t{ s_i
s_j }\right \rangle = \sum_{i,j=0}^t C_{ij},
\end{equation}
where the average is taken over different realizations and
$C_{ij}=\langle s_i s_j \rangle$ is the step-step correlation function.

Here the stationary, 
ergodic RWRW model differs crucially from the non-stationary
CTRW one, since for stationary RWRW processes
$C_{ij}$ only depends on the difference of its arguments
$C_{ij}=C_{|i-j|}$ and is an oscillating, slowly decaying function
(\textit{vide infra}). On the other hand, for 
CTRW different steps along the backbone of the comb are uncorrelated, so that
$C(i,j)= M(i) \delta_{i,j}$ with $M(i)$
being the probability that the time-step $t=i$ corresponds to a step on the
backbone and not on a tooth, i.e. to the rate of relevant steps. This one can be evaluated
via $\chi(n,t)$~\cite{KlafterSokolov} and for large $t$ behaves as 
$M(t) \simeq (2 \pi)^{-1/2} \; t^{-1/2}$. The explicit dependence of $C(i,j)$
on time is a fingerprint of the non-stationarity of CTRW. 

To asses $C_\tau =\left\langle s_n s_{n+\tau} \right\rangle$ for a RWRW we have to 
look separately at even and odd time lags $\tau = |i-j|$. The first step $s_n$
of the RW corresponds to zero delay and is taken to be in a 
positive direction along the track (to the right). Since the track of
the RW is an uncorrelated RW itself, the two steps are correlated only if they correspond 
to the motion over the same step (bond) of the track, so that $s_{n}s_{n+\tau} = \pm 1$ depending whether
the bond is crossed in the same direction (at even lags) or in the opposite directions (at odd lags), and are uncorrelated otherwise. 
Crossing of the same bond takes place with probability 1/2 and only if the walker 
was at its starting position (even lags) or at its position after performing the first step
(odd lags). For $\tau=2k$ the probability that the RW on a track returns to its starting point 
is given by the binomial distribution $\mathrm{binom}(k,2k-1,1/2)$ (where
$\mathrm{binom}\left(k,n,p\right)=\frac{n!}{k! (n-k)!}
p^k(1-p)^{n-k}$). For $\tau = 2k+1$ the return to the endpoint of  
the first step takes place with the probability
$\mathrm{binom}(k, 2k, 1/2)$.
We therefore have $C_0 = 1$ and $ C_1 = -1/2$, and generally:
\begin{eqnarray}\label{eq:Cn}
&& C_{2k} = (1/2) \mathrm{binom}\left(k,2k-1,1/2\right) \nonumber \\
&& C_{2k+1} = (-1/2)  \mathrm{binom}\left(k,2k,1/2\right).
\end{eqnarray}
Thus a RWRW model is a bona fide anti-persistent RW
model. Eq.~(\ref{eq:Cn})  has been numerically verified.

Note that both correlation functions, the one for RWRW and the one for
the comb, lead to the same MSDs. 
For RWRW we have:
$\left\langle x^2_n \right\rangle = 2  \sum_{k=0}^n (n-k) C_k - C_0 -C_n$.
For large $t=n$ the two last
contributions $C_0 =1$ and $C_n \to 0$ can be neglected compared to the sum and $C_k$ can be approximated via the
Stirling formula giving $C_k = \pm(2\pi k)^{-1/2}$. Passing from the sum to the integral 
we get for large $n$ $\left\langle x^2_n \right\rangle = \sqrt{2\pi^{-1}} n^{1/2}$.
For the comb we get
$\left\langle x^2_n \right\rangle = \sum_{k=1}^n (2\pi)^{-1/2} \; k^{-1/2} \approx \int_0^n \sqrt{2/\pi} \; k^{-1/2} dk = \sqrt{2\pi^{-1}} n^{1/2}$,
same as above.

Let us summarize our findings. We discussed the behavior of two popular
models of geometrically induced subdiffusion: the random walk on a comb (the model 
of the CTRW class) and the random walk on the random walk's trajectory, belonging
to the class of anti-persistent random walks. Although both models lead to
\textit{exactly the same} PDFs at all times, they
are crucially different with any other respect like FPTs or
correlation functions, which mirror the fact that the RWRW model is stationary
and the CTRW model is not. 

The difference in these behaviors leads us to a following conclusion.
We say that a model is described by some equation if several
relevant properties of the model can be obtained by solving
this equation with corresponding initial and boundary
conditions. For example, the fractional diffusion equation 
describes the comb model because not only the Green's function (PDF $p(x,t)$, the solution for
free boundaries), but also the FPT distribution to the point $x_1$ follows from the corresponding equation, 
now with the boundary condition $p(x_1,t)=0$. This is definitely not the case
for the RWRW model, for which we have to admit that the knowledge of the
Green's function (and of the equation for it) is not enough to obtain the FPT, and the model
is not fully described by this equation. Any conclusions about the nature
of the process based on the PDF (or moments) alone or on the knowledge of the equation
for such a PDF may thus stand on a quite shaky basis.

\section{Supplementary Material}

\subsection{MSD of the comb model}
Let us define the total number of time steps as
$n$ and the number of steps taken in the $x$-direction as $n_x$. The
number of steps taken in the $y$-direction is then $n_y = n - n_x$. Using CTRW
arguments~\cite{Metzler2000} and given that the distribution of
soujourn times in the teeth is given by $\psi(t) \simeq
\frac{1}{\sqrt{2 \pi}} t^{-3/2}$,
the one-dimensional ensemble MSD along the backbone takes the form:
\begin{equation}\label{eq:msdx}
\langle x^2\rangle_{ens} \sim \langle n_x\rangle \sim n^{1/2}.
\end{equation}
Similarly, for the $y$-axis we have: 
\begin{equation}\label{eq:msdy}
\langle y^2 \rangle_{ens} \sim \langle n_y \rangle = \langle n - n_x
\rangle = n - n^{1/2}.
\end{equation}
The two-dimensional MSD is then the sum of the two one-dimensional
MSDs, resulting in a linear time dependence:
\begin{equation}\label{eq:msdr}
\langle r^2 \rangle = \langle x^2 \rangle  + \langle y^2 \rangle = n.
\end{equation}
This result is numerically verified, displayed in the inset of Fig.~3
in the main text.

One should note that one commonly  treats the motions in
the $x$ and $y$ directions as independent~\cite{Sokolov2010a, Weiss} which leads to
a two-dimensional propagator being the product of the
two one-dimensional propagators. This, in turn, leads to a 
two-dimensional MSD non-linear in time, the one behaving as 
\begin{equation}
\langle r^2 \rangle_{ens} = \langle x^2 \rangle_{ens} + \langle y^2
\rangle_{ens} \sim n^{1/2} + n.
\end{equation}
Although this is asymptotically linear, the numerics in
Fig.~3 from the main text clearly exhibit a completely linear time
dependence at all time scales including the shortest ones, thus
rejecting this view in favour of the one presented above.

Eq.~\ref{eq:msdr} describes \textit{normal diffusion} on the comb
structure, which, contrary to the diffusion in the 
projection on the $x$-direction, is a stationary and ergodic random process.
We also note that a random walk on a percolation cluster at criticality is
subdiffusive, meaning that the comb structure is inherently inadequate
to model a percolation cluster. 

\subsection{The FPT distributions}

Let us obtain the FPT distribution of the RWRW model, again by averaging over the initial
point and the realizations of the walk. 
We fix the starting point $x=0$ and the final point $x_1$. The RW trajectory on which the
random walk than takes place has two strands connecting these two
points (since in 1D a RW is certain to reach any given point on a line), and therefore the
corresponding random walk takes place on a finite interval. The distances $l_1$ and $l_2$ along
the corresponding strands are independent random variables, both distributed according to the Smirnov law
\begin{equation}\label{eq:smirnov} 
f(l_{1,2}| x) = \frac{x}{2 \sqrt{\pi} l_{1,2}^{3/2}} \exp\left(- \frac{x^2}{4 l_{1,2}}\right). 
\end{equation}
Therefore the corresponding RW on the RW takes place at an interval $(-l_1,l_2)$ starting at
$l=0$, or, equivalently, on an interval of length $L=l_1+l_2$ starting at a point at a distance
$l_1$ from the left boundary of the interval. 
It is convenient to change the variables, and to consider a symmetric interval $(-L/2,L/2)$
with the starting point  $y = (l_1-l_2)/2$ inside it (being at the distance $l_1$ from its left boundary
and at the distance $l_2$ from the right one). 
The probability to find a particle at position $l$ within this interval with
absorbing boundaries (in the diffusion approximation, $D=1/2$) is given by 
\begin{equation}
P(l,t) = \frac{2}{L}\sum_{m=0}^{\infty} \cos\left(b_m
  l\right)\cos\left(b_m y\right) \exp{( -b_m^2 Dt)},
\end{equation}
where $b_m = \pi(2m+1)/L$.
Integration over $l$ gives us the survival probability
\begin{equation}
\Phi_{1,2}(t) = \sum_{m=0}^{\infty} (-1)^m \frac{4}{L b_m}
\cos\left(b_m y\right)
\exp{( -b_m^2 Dt)}
\end{equation}
which has to be averaged over the position of the starting point $y$ and of the length of the interval.
The overall explicit evaluation of the mean of this series is tedious, but finding the long time asymptotic behavior
is relatively simple. To do this, we note that at long times only the first term ($m=0$) plays a role, so that
\begin{equation}\label{eq:phi12}
\Phi_{1,2}(t) \approx  \frac{4}{\pi} \cos\left(\frac{\pi y}{2L}\right)
\exp{\left( -\frac{\pi^2}{L^2} Dt \right)},
\end{equation}
where $Dt=n/2$.
The average over the position of the starting point $y$ and of the length of the interval is given by:
\begin{equation}\label{eq:phi_int}
\Phi(t) = \iint\limits^{\:\:\:\:\:\:\:\:\infty}_0\Phi_{1,2}(t) p(l_1)p(l_2) \mathrm{d}l_1 \mathrm{d}l_2.
\end{equation}
We substitute Eqs.~(\ref{eq:phi12}) and~(\ref{eq:smirnov}) in
Eq.~(\ref{eq:phi_int}) and make a change of variables, $L = l_1 + l_2$
and $z = l_2/l_1$ (the Jacobian of this transformation is $
\partial(l_1,l_2)/\partial(L,z)= L/(1+z)^2$). Making a
further transformation, $\xi = 1/L$, leads to 
\begin{equation}
\Phi(t) = \frac{x^2}{\pi^2}\iint\limits^{\:\:\:\:\:\:\:\:\infty}_0
\frac{1+z}{z^{3/2}} \cos(\frac{\pi}{2}\frac{1-z}{1+z})
e^{( -\pi^2 Dt \xi^2 -\frac{x^2(1+z)^2}{4z} \xi )} \mathrm{d}\xi \mathrm{d}z.
\end{equation} 

We are interested in the asymptotic behavior of this function for $t \to \infty$. The integral over $\xi$ can be taken in quadratures:
\begin{eqnarray*}
I_1 &=& \int_0^\infty \exp\left( -\pi^2 Dt \xi^2 -\frac{x^2(1+z)^2}{4z} \xi \right) d\xi \\
&=& \frac{1}{2\sqrt{\pi Dt}} \exp \left(\frac{x^4(1+z)^4}{64 z^2 \pi^2 Dt}\right) 
\mathrm{erfc}\left(\frac{x^2(1+z)^2}{8 z \pi \sqrt{Dt}}\right)
\end{eqnarray*}
and tends to 
\begin{equation}
I_1 \to \frac{1}{2\sqrt{\pi Dt}}
\end{equation}
for $t \to \infty$ (unless $z$ gets extremely large, when it starts to decay as $z^{-1}$). Such large values of 
$z$ do not play any role, since the integral 
\begin{equation}
I_2 = \int_0^\infty \cos\left(\frac{\pi}{2}\frac{1-z}{1+z}\right)
\frac{1+z}{z^{3/2}}dz
\end{equation}
converges: It is equal to  $\,_1 F_2\left(1,\frac{1}{2},\frac{3}{2},-\frac{\pi^2}{16} \right)$ 
where $ \,_1 F_2$ is the generalized hypergeometric function (the integral can be 
transformed to the form (Eq. 2.5.8.1 in~\cite{Prudnikov}) by the change of variable $y=(1-z)/(1+z)$ ). The numerical evaluation of the
integral gives approximately the value $I_2 = 11.2$. This numerical evaluation results in the
survival
probability whose asymptotical behaviour at $t\rightarrow \infty$
follows: 
\begin{equation}
\Phi(t) \simeq C \frac{x^2}{\sqrt{Dt} },
\end{equation}
where $C = I_2/(2 \pi^{5/2}) \approx 0.64$.

\subsection{Marginal distributions for the RWRW}
For the sake of completeness we give here the expression for $P(L,z)$,
the joint probability density that a point on a track chosen at random
falls in the interval of length $L$ between its two crossing points
with the $x=const$ line and divides this interval into two parts whose
length quotient is $l_1/l_2 = z$. This one is obtained by the variable transformation
from $p(l_1,l_2)=p(l_1|x)p(l_2|x)$ with $p(l|x)$ given by Eq.~(\ref{eq:smirnov}) and reads
\begin{equation}
P(L,z)= \frac{x^2}{4 \pi} 
\frac{1+z}{z^{3/2}}\frac{1}{L^2} \exp\left(-\frac{(1+z)^2}{z}\frac{x^2}{4L} \right).
\end{equation}
The marginal distribution in $z$ follows elementary~\cite{Sokolov2010}:
\begin{equation}
p(z)=\int_0^\infty P(L,z)dL = \frac{1}{\pi} \frac{1}{\sqrt{z}(1+z)}.
\end{equation}
This distribution corresponds to a very uneven distribution of the starting point
within the interval (the equidistribution would correspond to 
$p(z)=(1+z)^{-2}$, and doesn't have a singularity at small $z$). This means, that in our case the particle starts close to
one of the ends of the interval (the distributions of $z$ and $1/z$ are the same), and therefore
the FPT distribution decays faster than in the case of the equidistribution.

To calculate the marginal distribution in $L$, one 
first passes to
the distribution of the variable $\xi=1/L$:
\begin{equation}
P(\xi,z)= \frac{x^2}{4 \pi} 
\frac{1+z}{z^{3/2}} \exp\left(-\frac{(1+z)^2}{z}\frac{x^2}{4}\xi \right),
\end{equation} 
takes the Laplace transform in $\xi$ to obtain
\begin{equation}
\tilde{P}(u,z) = \frac{1}{\pi}\frac{x^2(1+z)}{\sqrt{z}(4uz+x^2(1+z)^2)},
\end{equation}
performs integration in $z$ (which is an elementary but tedious procedure) to obtain
\begin{equation}
\tilde{p}(u) = \int_0^\infty  \frac{1}{\pi}\frac{x^2(1+z)}{\sqrt{z}(4uz+x^2(1+z)^2)}dz = \frac{x}{\sqrt{u+x^2}}
\end{equation}
takes the inverse Laplace transform,
\begin{equation}
p(\xi)=\frac{x}{\sqrt{\pi \xi}} \exp(-x^2 \xi)
\end{equation}
and finally returns to $L$ to get
\begin{equation}
p(L)=\frac{1}{\sqrt{\pi}L^{3/2}}\exp\left(-\frac{x^2}{L} \right),
\end{equation}
the Smirnov distribution, as anticipated.

Note that neglecting the role of the initial position (which, as stated, is extremely unevenly distributed within
the $L$-interval) would lead to a different result, namely to:
\begin{equation}
\Phi(t) 
= \frac{x}{2 \pi^{3/2}\sqrt{Dt}}
e^{\left( \frac{x^4}{8 \pi^2 Dt} \right)} K_{1/4}\left( \frac{x^4}{8 \pi^2 Dt} \right)
\end{equation}
(see Eq. 2.3.15.5 in ~\cite{Prudnikov}). Using the behavior of the modified Bessel functions for small values of its
argument, $K_{1/4}(z) \simeq \frac{1}{2} \Gamma(1/4) (z/2)^{-1/4}$ (see Eq. 9.6.9 in~\cite{Abramovitz}) we 
would get that $\Phi(t) \propto  t^{-1/4}$ like in the CTRW.


\begin{thebibliography}{15}
\expandafter\ifx\csname natexlab\endcsname\relax\def\natexlab#1{#1}\fi
\expandafter\ifx\csname bibnamefont\endcsname\relax
  \def\bibnamefont#1{#1}\fi
\expandafter\ifx\csname bibfnamefont\endcsname\relax
  \def\bibfnamefont#1{#1}\fi
\expandafter\ifx\csname citenamefont\endcsname\relax
  \def\citenamefont#1{#1}\fi
\expandafter\ifx\csname url\endcsname\relax
  \def\url#1{\texttt{#1}}\fi
\expandafter\ifx\csname urlprefix\endcsname\relax\def\urlprefix{URL }\fi
\providecommand{\bibinfo}[2]{#2}
\providecommand{\eprint}[2][]{\url{#2}}

\bibitem[{\citenamefont{Weiss and Havlin}(1986)}]{Weiss1986}
\bibinfo{author}{\bibfnamefont{G.~H.} \bibnamefont{Weiss}} \bibnamefont{and}
  \bibinfo{author}{\bibfnamefont{S.}~\bibnamefont{Havlin}},
  \bibinfo{journal}{Physica A} \textbf{\bibinfo{volume}{134}},
  \bibinfo{pages}{474} (\bibinfo{year}{1986}).

\bibitem[{\citenamefont{Kehr and Kutner}(1982)}]{Kehr1982}
\bibinfo{author}{\bibfnamefont{K.~W.} \bibnamefont{Kehr}} \bibnamefont{and}
  \bibinfo{author}{\bibfnamefont{R.}~\bibnamefont{Kutner}},
  \bibinfo{journal}{Physica A} \textbf{\bibinfo{volume}{110}},
  \bibinfo{pages}{535} (\bibinfo{year}{1982}).

\bibitem[{\citenamefont{Doi and Edwards}(1988)}]{Doi1988}
\bibinfo{author}{\bibfnamefont{M.}~\bibnamefont{Doi}} \bibnamefont{and}
  \bibinfo{author}{\bibfnamefont{S.~F.} \bibnamefont{Edwards}},
  \emph{\bibinfo{title}{The Theory of Polymer Dynamics}}
  (\bibinfo{publisher}{Oxford Science Press}, \bibinfo{year}{1988}).

\bibitem[{\citenamefont{Metzler and Klafter}(2000)}]{Metzler2000}
\bibinfo{author}{\bibfnamefont{R.}~\bibnamefont{Metzler}} \bibnamefont{and}
  \bibinfo{author}{\bibfnamefont{J.}~\bibnamefont{Klafter}},
  \bibinfo{journal}{Physics Reports} \textbf{\bibinfo{volume}{339}},
  \bibinfo{pages}{1} (\bibinfo{year}{2000}).

\bibitem[{\citenamefont{Chandrasekhar}(1943)}]{Chandrasekhar1943}
\bibinfo{author}{\bibfnamefont{S.}~\bibnamefont{Chandrasekhar}},
  \bibinfo{journal}{Rev. Mod. Phys.} \textbf{\bibinfo{volume}{15}},
  \bibinfo{pages}{89} (\bibinfo{year}{1943}).

\bibitem[{\citenamefont{Klafter and Sokolov}(2011)}]{KlafterSokolov}
\bibinfo{author}{\bibfnamefont{J.}~\bibnamefont{Klafter}} \bibnamefont{and}
  \bibinfo{author}{\bibfnamefont{I.~M.} \bibnamefont{Sokolov}},
  \emph{\bibinfo{title}{First Steps in Random Walks}}
  (\bibinfo{publisher}{Oxford University Press, Oxford}, \bibinfo{year}{2011}).

\bibitem[{\citenamefont{Sokolov
  et~al.}(2001{\natexlab{a}})\citenamefont{Sokolov, Blumen, and
  Klafter}}]{Sokolov2001}
\bibinfo{author}{\bibfnamefont{I.~M.} \bibnamefont{Sokolov}},
  \bibinfo{author}{\bibfnamefont{A.}~\bibnamefont{Blumen}}, \bibnamefont{and}
  \bibinfo{author}{\bibfnamefont{J.}~\bibnamefont{Klafter}},
  \bibinfo{journal}{Europhysics Letters} \textbf{\bibinfo{volume}{56}},
  \bibinfo{pages}{175} (\bibinfo{year}{2001}{\natexlab{a}}).

\bibitem[{\citenamefont{Sokolov
  et~al.}(2001{\natexlab{b}})\citenamefont{Sokolov, Blumen, and
  Klafter}}]{Sokolov2001a}
\bibinfo{author}{\bibfnamefont{I.~M.} \bibnamefont{Sokolov}},
  \bibinfo{author}{\bibfnamefont{A.}~\bibnamefont{Blumen}}, \bibnamefont{and}
  \bibinfo{author}{\bibfnamefont{J.}~\bibnamefont{Klafter}},
  \bibinfo{journal}{Physica A: Statistical Mechanics and its Applications}
  \textbf{\bibinfo{volume}{302}}, \bibinfo{pages}{268}
  (\bibinfo{year}{2001}{\natexlab{b}}).

\bibitem[{Sup()}]{SupMat1}
\bibinfo{note}{For a detailed calculation we refer to the Supplementary
  Material.}

\bibitem[{\citenamefont{Meroz et~al.}(2011)\citenamefont{Meroz, Sokolov, and
  Klafter}}]{Meroz2011}
\bibinfo{author}{\bibfnamefont{Y.}~\bibnamefont{Meroz}},
  \bibinfo{author}{\bibfnamefont{I.~M.} \bibnamefont{Sokolov}},
  \bibnamefont{and} \bibinfo{author}{\bibfnamefont{J.}~\bibnamefont{Klafter}},
  \bibinfo{journal}{Phys. Rev. E} \textbf{\bibinfo{volume}{83}},
  \bibinfo{pages}{020104} (\bibinfo{year}{2011}).

\bibitem[{\citenamefont{Sokolov and
  Eliazar}(2010{\natexlab{a}})}]{Sokolov2010a}
\bibinfo{author}{\bibfnamefont{I.~M.} \bibnamefont{Sokolov}} \bibnamefont{and}
  \bibinfo{author}{\bibfnamefont{I.~I.} \bibnamefont{Eliazar}},
  \bibinfo{journal}{Phys. Rev. E} \textbf{\bibinfo{volume}{81}},
  \bibinfo{pages}{026107} (\bibinfo{year}{2010}{\natexlab{a}}).

\bibitem[{\citenamefont{Weiss}(1994)}]{Weiss}
\bibinfo{author}{\bibfnamefont{G.~H.} \bibnamefont{Weiss}},
  \emph{\bibinfo{title}{Aspects and Applications of the Random Walk}}
  (\bibinfo{publisher}{North Holland Press, Amsterdam}, \bibinfo{year}{1994}).

\bibitem[{\citenamefont{Prudnikov et~al.}(1988)\citenamefont{Prudnikov,
  Brychkov, and Marichev}}]{Prudnikov}
\bibinfo{author}{\bibfnamefont{A.}~\bibnamefont{Prudnikov}},
  \bibinfo{author}{\bibfnamefont{Y.}~\bibnamefont{Brychkov}}, \bibnamefont{and}
  \bibinfo{author}{\bibfnamefont{O.}~\bibnamefont{Marichev}},
  \emph{\bibinfo{title}{Integrals and Series}} (\bibinfo{publisher}{Gordon \&
  Breach Science Publishers/CRC Press}, \bibinfo{year}{1988}).

\bibitem[{\citenamefont{Sokolov and Eliazar}(2010{\natexlab{b}})}]{Sokolov2010}
\bibinfo{author}{\bibfnamefont{I.}~\bibnamefont{Sokolov}} \bibnamefont{and}
  \bibinfo{author}{\bibfnamefont{I.}~\bibnamefont{Eliazar}},
  \bibinfo{journal}{Phys. Rev. E} \textbf{\bibinfo{volume}{81}},
  \bibinfo{pages}{026107} (\bibinfo{year}{2010}{\natexlab{b}}).

\bibitem[{\citenamefont{Abramovitz and Stegun}(1964)}]{Abramovitz}
\bibinfo{author}{\bibfnamefont{M.}~\bibnamefont{Abramovitz}} \bibnamefont{and}
  \bibinfo{author}{\bibfnamefont{I.}~\bibnamefont{Stegun}},
  \emph{\bibinfo{title}{Handbook of Mathematical Functions}}
  (\bibinfo{publisher}{Dover}, \bibinfo{year}{1964}).

\end{thebibliography}
\end{document}